\documentclass[aps,prl,twocolumn,groupedaddress,showpacs]{revtex4}
\usepackage[]{epsfig}
\newcommand{\la}{\langle}
\newcommand{\ra}{\rangle_L} 
\newcommand{\tr}{t_{\rm rel}}

\begin{document}

\title{Finite size scaling analysis of the glass transition}

\author{Ludovic Berthier}
\altaffiliation{Also at:
Laboratoire des Verres, Universit\'e Montpellier II,
34095 Montpellier, France}
\affiliation{Theoretical Physics, 1 Keble Road, Oxford, OX1 3NP, UK}

\date{\today}

\begin{abstract}
We show that finite size scaling techniques
can be employed to study the glass transition.
Our results follow from the postulate of
a diverging correlation length at the glass transition 
whose physical manifestation is the presence of dynamical heterogeneities.
We introduce a parameter $B(T,L)$ whose temperature, $T$, and 
system size, $L$, dependences permit a precise location of 
the glass transition. We discuss the finite size scaling
behaviour of a diverging susceptibility $\chi(L,T)$.
These new techniques are successfully used to study two lattice models.
The analysis straightforwardly applies to any glass-forming system.
\end{abstract}



\maketitle

Progress in the study of the `glass transition' has
been slow for two main reasons. First, 
almost by definition, the dynamics of 
a liquid supercooled through its melting transition towards
its glass phase becomes so slow that
it is impossible 
to actually cross a possible phase transition while staying 
at equilibrium. One relies therefore 
on thermodynamic or kinetic extrapolations
which stop at the glass temperature $T_g$ where 
the system no longer equilibrates.
As a consequence, the mere existence of a genuine
transition
is still questioned, even though experiments now 
probe more than a decade of decades 
of relaxation times.

Secondly, the nature of the putative transition remains largely unknown as
compared to more standard phase transitions. Basically, the situation
is such that the formulation of a Ginzburg-Landau type of model 
is still impossible, because one does not know what would be the correct 
order parameter, its symmetries, dynamics, correlations, etc. 
Similarly, at the beginning of the 90's, two papers 
apparently settled the 
related question of the possible existence 
of a diverging correlation length
at the glass transition, reporting the absence of such a length 
scale~\cite{nagel,dasgupta}.
As a regrettable corollary, modern developments 
of statistical mechanics such as 
renormalization group, finite size scaling, or universality classes
became apparently useless to study the glass transition.

In this paper, we show that finite size scaling techniques
first derived from renormalization group concepts
in the context of continuous phase transitions
are useful to study the glass transition, at variance with
the common belief.
This should allow much more powerful extrapolations to locate
a possible transition and a better characterization of
its nature, thus making progress 
on the two important points mentioned above.

The paper is organized as follows. 
We first describe the observations
underlying our work, before 
describing the principles of the analysis.  
As a first application of the method, 
we then consider two well-known models proposed 
in the context of the glass transition, namely the one-dimensional
Fredrickson-Andersen model~\cite{FA} and the three-dimensional 
Kob-Andersen model~\cite{KA}.

Our work is based upon the recent burst of activity related
to the definition, observation and characterization of 
dynamical heterogeneities in the dynamics of supercooled 
liquids~\cite{reviewexp}.
The underlying physics is that particles
move in a cooperative manner, 
the more so the nearer the glass transition. 
This picture immediately suggests 
the existence of a
correlation length $\xi(T,t)$ associated to
these dynamical domains, reflecting the length
scale on which the particles dynamics
have been correlated in a time interval $t$ 
at temperature $T$.
We define a `coherence length' $\ell(T) \equiv \xi(T,\tr)$, where
$\tr= \tr(T)$ is the mean relaxation time of the liquid.
Whether or not $\ell(T)$  is connected to some yet 
undiscovered structural length scale is an 
important open problem. 
Crucially, however, progress can be made if one postulates 
that $\ell(T)$ actually diverges at 
some---possibly zero---temperature $T_c$
which will thus non-ambiguously define the glass
transition temperature one seeks to determine.
It should coincide with the temperature at which
$\tr(T)$ also diverges. 

A measure of $\xi(T,t)$ in a liquid can be 
obtained via the spatial decay of the 
following correlation function:
\begin{equation}
C (r,t) = \la
F(r+r_0,t) F(r_0,t) \rangle  - \la F(r+r_0,t) \rangle
\la F(r_0,t) \rangle,
\label{bub}
\end{equation}
where $F(r,t) = \rho (r, t+t_0) \rho (r, t_0)$; 
$\rho (r,t)$ is the density at time $t$ and position $r$,
$r_0$ and $t_0$ are arbitrary position and time, respectively.
Alternatively, $F(r,t)$ can be replaced by any 
two-time function, $F(r,t) = A(r,t+t_0) B(r,t_0)$, where
$A$ and $B$ are physical observables.
In essence, Eq.~(\ref{bub}) is a two-time, 
two-point correlation function which measures
correlations in trajectory space~\cite{glo,jp}.
Note that the commonly discussed `non-Gaussian parameter' is related to
Eq.~(\ref{bub}) when $A$ and $B$ 
refer to the particle positions, but for $r=0$.
As such, it contains no information about $\ell(T)$.
In fact, Eq.~(\ref{bub}) 
captures most of the measurements of dynamical
heterogeneities which amounts to first
distinguish between `fast' and `slow' 
particles and then to check how these are spatially 
correlated~\cite{reviewexp}. 
The correlators $F(r,t)$ precisely tell us how fast a particle is, 
while Eq.~(\ref{bub}) measure the corresponding spatial
correlations.
The use of the dynamic function $F(r,t)$ makes an
arbitrary criterion to distinguish between particles 
unnecessary.
Variations of the correlator (\ref{bub})
have been studied. 
In Refs.~\cite{glo,parisi}, the density fluctuations were replaced 
by an arbitrarily defined overlap between configurations, and 
by  particle displacements in Refs.~\cite{basch,heuer,hiwa}.
These works show that $\xi(T,t)$ is 
indeed maximum for $t \approx \tr(T)$, which justifies
our definition of $\ell(T)$. Also, 
$\ell(T)$ decreases with $T$, although simulations are
yet too limited to confirm a possible divergence
at low temperature.

The observation of a growing coherence length scale 
has deep consequences which we start 
to explore in this paper.
This indeed suggests 
that local dynamical functions $F(r,t)$ become long-range correlated
when the glass phase is approached. 
As such they play a role similar to that of the order
parameter in a standard continuous phase transition.
This simple remark is central in our approach.
Having identified a quantity
analogous to an order parameter, it becomes a simple task 
to extend the tools developed for conventional transitions 
to study the problem of the glass transition, starting
here with finite size scaling techniques.

Define first the parameter $B(T,L)$ as
\begin{equation}
B(T,L) = 1 - \frac{\la \varphi^4 \ra - 4 \la \varphi^3 \ra \la \varphi \ra +
6 \la \varphi^2 \ra \la \varphi \ra^2 - 3 \la \varphi \ra^3}{3 
\big[ \la \varphi^2 \ra^2 
- 2 \la \varphi^2 \ra \la \varphi \ra^2 + \la \varphi \ra^4
\big] },
\label{BTL}
\end{equation}
with $\varphi=\varphi(\tr(T)) \equiv L^{-d} \int d^dr F(r,\tr(T))$, where
$d$ is the space dimensionality. 
The crucial information 
is that the average
$\la \cdots \ra$ is performed in a system of finite linear size $L$. 
By construction, $B(T,L)=0$ when the fluctuations of $\varphi$
are Gaussian, since it is built from the 
fourth cumulant of the probability distribution
function of $\varphi$, $P_L(\varphi)$.
This distribution is Gaussian when
$L \gg \ell(T)$ as a consequence of the central limit theorem.
However, $P_L(\varphi)$ becomes non-Gaussian when $L < \ell(T)$, and
$B(T,L)$ is therefore non-zero in that case. 
Moreover, it is natural to expect that 
$B(T,L)$ becomes a scaling function of the variable 
$\ell(T) / L$ in the regime $\ell(T), L \gg a$, 
where $a$ is the microscopic length scale of the problem, e.g.
the particle size in a simple liquid.
Also, $P_L(\varphi)$ should be independent of $L$ and 
$\ell(T)$ in the regime $a \ll L \ll \ell(T)$.
The latter assumption implies in particular 
that $B(T_c,L)$ is a constant `universal' number
since $L \ll \ell(T_c) = \infty$ is automatically satisfied
at the transition. Therefore, the curves $B(T,L)$ versus 
$T$ for various $L$ cross at $T_c$.
The definition (\ref{BTL}) is naturally inspired by the classical
paper~\cite{binder}, 
where a numerical finite size scaling analysis of the Ising model was
performed. In that case, $\varphi$ was the order
parameter of the paramagnetic-ferromagnetic transition, i.e.
the magnetization density.

We also define the susceptibility
\begin{equation}
\chi(T,L) = \frac{L^d}{T} \bigg[ \la \varphi^2 \ra - \la \varphi \ra^2 
\bigg],
\label{sus}
\end{equation}
again explicitly retaining its $L$ dependence. 
Similarly, one expects this function to 
exhibit a scaling behaviour for large 
$L$ and $\ell(T)$. 
When the correlator $C(r,t)$ is itself
a simple function of the ratio $r/\ell(T)$, one gets
\begin{equation}
\chi(L,T) \approx L^d \tilde{\chi} \left( 
\frac{\ell(T)}{L} \right),
\label{chiscal}
\end{equation}
where the scaling function $\tilde{\chi}(x)$ behaves 
as $\tilde{\chi} (x \ll 1) \approx x^d$ and
$\tilde{\chi}(x \gg 1) \approx const$. 
The assumption made on the correlator is supported by 
all known simulation results so far~\cite{glo,parisi,heuer}. It is 
satisfied also in the two models studied below. 
Using with caution the language of critical phenomena, this would
mean that the anomalous exponent $\eta$ is 
$\eta = 2-d$, implying a simple Fisher's law
$\gamma = \nu d$ between the exponents
of the susceptibility and the correlation length.
This point certainly needs further investigation
in realistic supercooled liquids.

The suggested analysis requires 
extrapolations towards the glass phase, as all 
equilibrium measurements in supercooled liquids do. However, 
the new quantities $B(T,L)$ and $\chi(T,L)$ defined above are much richer
quantities than, say, an average relaxation time since they 
contain information on the full distribution $P_L(\varphi)$.
This will be clearly demonstrated in the rest of the paper.

\begin{figure}
\psfig{file=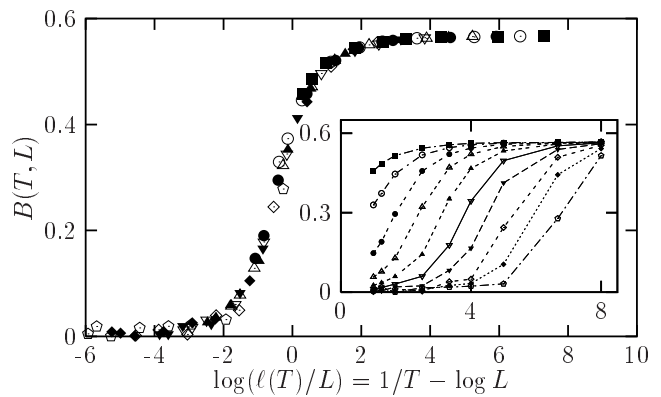,width=8.65cm}
\caption{\label{fa}
Inset: The parameter $B(T,L)$ as a function of 
the inverse temperature $1/T$
for various system sizes $L=2$, 4, $\cdots$, 1024 (top to bottom)
cross at the glass transition $T_c=0$.
Main: The parameter $B(T,L)$ as a function
of the scaling variable $\log ( \ell (T)/ L )$ with
$\ell (T)= \exp (1/T)$.
Both figures are for the one-dimensional spin facilitated Ising model.}
\end{figure}

We now use the analysis suggested above 
to study two lattice models. We start with the 
one-dimensional version of the spin 
facilitated Ising model introduced in Ref.~\cite{FA}.
Spin facilitated models have been introduced as caricatures
of real supercooled liquids, in the sense that they are
non-disordered 
spin models displaying glassy dynamics with only very simple 
static correlations.
We have chosen this version of the model since it is reasonably well 
characterized~\cite{jp,FA,harrowell}. 
Moreover, as a one-dimensional lattice model, 
it is relatively easy to simulate on very large time and length scales, 
nicely confirming the expected scaling behaviour. These results thus 
constitutes a very interesting first application 
of the methods described above.
The model is defined by the Hamiltonian
\begin{equation}
H =  \sum_{i=1}^N  s_i,
\label{hamfa}
\end{equation}
where $s_i=0,1$ are two-state observables located 
at the sites of
a chain of size $N$ with periodic boundary conditions. 
Glassiness originates from the chosen dynamics,
since the static properties of the 
non-interacting Hamiltonian (\ref{hamfa})
are completely trivial. 
In a standard Metropolis algorithm, the transition rates are
$w(s_i \to 1-s_i) = \min(1,e^{-\beta \Delta E})$, 
where $\Delta E$ is the energy change in the transition
$s_i \to 1-s_i$.
Here, the rates are given instead
by $w'(\Delta E) = ( s_{i-1} + s_{i+1} )
w(\Delta E)$ meaning that a spin can flip 
only if it has at least one nearest neighbor 
whose value is 1.
This kinetic constraint implies glassiness.
From (\ref{hamfa}), one has
$\la s_i \rangle = (1+e^{1/T})^{-1}$, so that
the density of spin 1 becomes small at low $T$ therefore
allowing less and less transitions. 
This leads to an Arrhenius relaxation time, $\tr = \exp(3/T)$, 
which makes the model `strong' according to the classification
commonly used in the glass literature.
Also, the dynamics becomes more and more 
cooperative when approaching the glass transition at $T_c = 0$, 
with a diverging coherence length 
$\ell(T) = \exp(1/T)$~\cite{jp,harrowell}.

\begin{figure}
\psfig{file=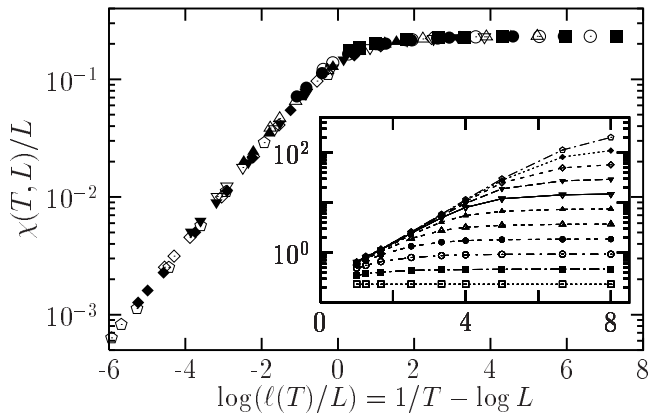,width=8.65cm}
\caption{\label{chi} Inset: The susceptibility $\chi(T,L)$
as a function of the inverse temperature $1/T$ for 
the same sizes as Fig.~\ref{fa} increasing from bottom to top.
Main: The rescaled susceptibility $\chi(T,L) L^{-1}$
as a function of the scaling variable $\log ( \ell (T)/ L )$ with
$\ell (T)= \exp (1/T)$.
Both figures are 
for the one-dimensional spin facilitated Ising model.}
\end{figure}

We simulate this model using a continuous time algorithm 
in the temperature interval
$T \in [0.125, 1.0]$, approximately corresponding to
13 decades in relaxation times. For each temperature, 
a large system of $N=2 \cdot 10^4$ spins is studied, and
averages are performed in subsystems of finite size
$L \in [2, 1024]$ for a number of 
independent configurations ranging from 200 to 20000.
Fortunately, more statistics is required at higher temperatures
where deviations from Gaussian behaviour are smaller.
A natural choice for the dynamic function $F(i,t)$ in
the definition (\ref{BTL}) is the persistence 
of the spin $s_i$ at site $i$.
The mean persistence is given by $\la F(i,t) \rangle_{L=\infty} = 
\exp \left[ - \sqrt{t/\tr(T)} \right]$, which defines
the relaxation time. 

Our results for this model
are presented in Figs.~\ref{fa} and \ref{chi}. 
The inset of Fig.~\ref{fa} shows the variation of $B(T,L)$
as a function of $1/T$ for various sizes $L$.
As expected, $B(T,L)$ crosses over from $B(T,L)=0$ at high $T$ to a non-zero
value at low $T$, the locus of the crossover being $L$-dependent.
The various curves clearly cross at $T_c=0$ only.
One observes in the
main figure an excellent collapse of the data, showing
that $B(T,L)$ satisfies the expected scaling form
$B(T,L) \approx \tilde{B} (\ell(T) /L)$. Since
$P_L(\varphi)$ is bimodal 
when $L \ll \ell(T)$, it is a simple task 
to compute $B(T_c=0,L) = \tilde{B} (x \to \infty) = 0.566...$.
The inset of Fig.~\ref{chi} presents the susceptibility $\chi(T,L)$
as a function of $1/T$ for various sizes $L$. Again, one clearly sees the
expected behaviours: $\chi(T,\infty)$ diverges when $T \to 0$ as 
$e^{1/T} = \ell(T)$, while for finite $L$ the divergence is 
smeared when $\ell(T) > L$. The main figure 
shows the perfect collapse of the data suggested by
Eq.~(\ref{chiscal}) when $\chi(L,T) L^{-1}$ is plotted
as a function of $\ell(T)/L$ (remember that $d=1$).

As a second application of the techniques described
above, we study the kinetically constrained lattice gas introduced
in Ref.~\cite{KA}. The model consists of hard spheres 
on a cubic lattice of linear size $N$.
In this case, temperature plays no role, 
and the relevant control parameter is the density
of particles, $\rho$.
This lattice gas is also complemented 
by kinetic rules which make the dynamics glassy. 
In a non-constrained lattice gas, particles move to an empty
nearest neighbour site with unit rate. 
In the present model, particles can move to an empty nearest neighbour, 
provided the particle has fewer than 4 neighbours before and after
the move. This kinetic rule aims at 
reproducing dynamics of a supercooled liquid,
where particles can hardly escape the cage formed by their neighbours.
As a simple lattice model for the glass transition, it has been 
much studied since its introduction.
Simulations have reported a dynamical arrest at a 
density $\rho_c \approx 0.881$ where the relaxation time
apparently diverges as a power law~\cite{KA}.
Very recently, however, a lower bound for the 
diffusion constant was analytically derived for this model, 
showing that the divergence of the relaxation time at a finite density
found in simulations is only apparent~\cite{giulio}.
The growth of $\tr(\rho)$ with $\rho$ is in fact
extremely abrupt, $\log \log \tr(\rho) \propto (1-\rho)^{-1}$,
with an associated coherence length $\log \log \ell(\rho) = c/(1-\rho)$, 
where $c$ is a numerical constant.
The extremely fast increase of $\tr$ can easily be 
numerically confused
with a true divergence at a finite density.
This more difficult problem represents thus a highly selective test for
the methods described in this paper. We now show 
that finite size scaling allows one
to distinguish between a true and an apparent
divergence of the relaxation time in this model. 

We use again a continuous time Monte Carlo algorithm to study 
a model of linear size $N=24$ for densities $\rho \in [0.75,0.87]$,
covering almost 8 decades of relaxation times. 
We use $F(i,t) = \rho(i,t) \rho(i,0)$ as a dynamical quantity, 
where $\rho(i,t)$ is the density at site $i$ and time $t$.
The averages are performed in subsystems of linear sizes $L$, 
using $10^4$ independent initial conditions at each density.
The correlator (\ref{bub}) is well described by a simple 
exponential form, $C(r,\tr) = e^{-r/\ell(\rho)}$. The latter measurements 
are also a check against spurious finite size effects, since they
show that $N \gg \ell(\rho)$ for the densities studied here.

\begin{figure}
\psfig{file=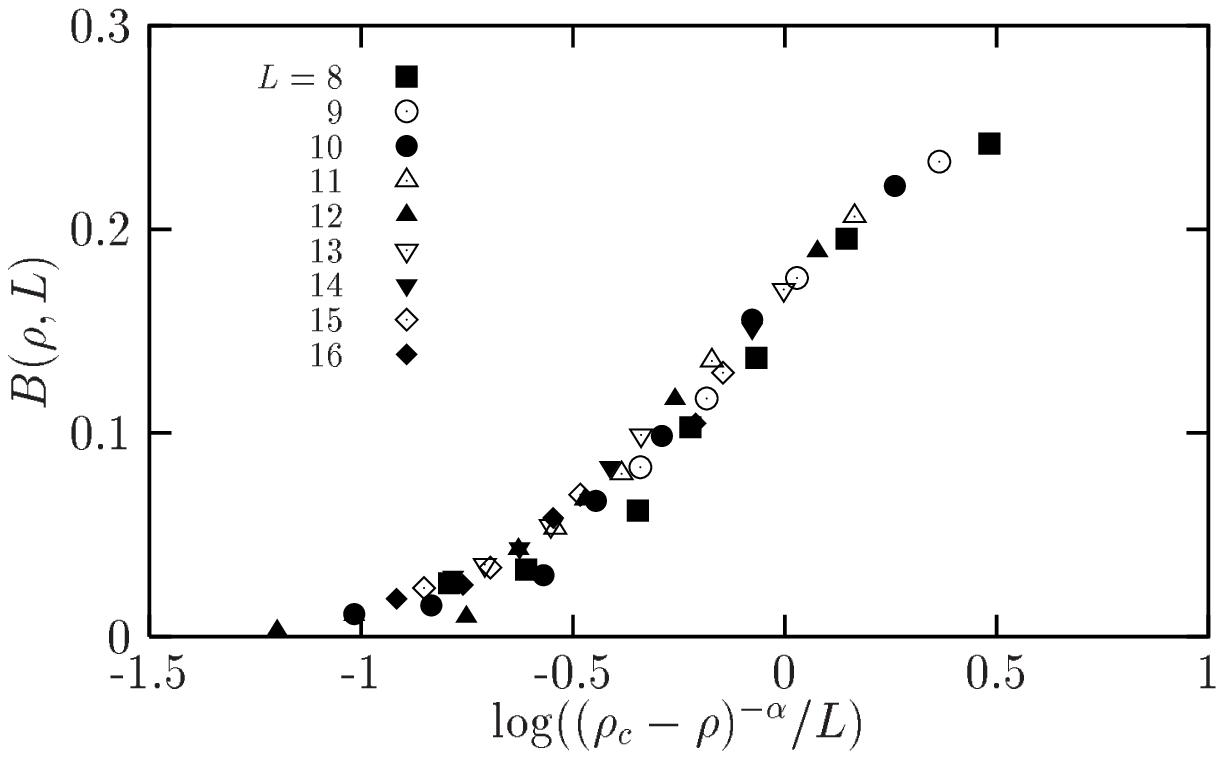,width=8.65cm}
\psfig{file=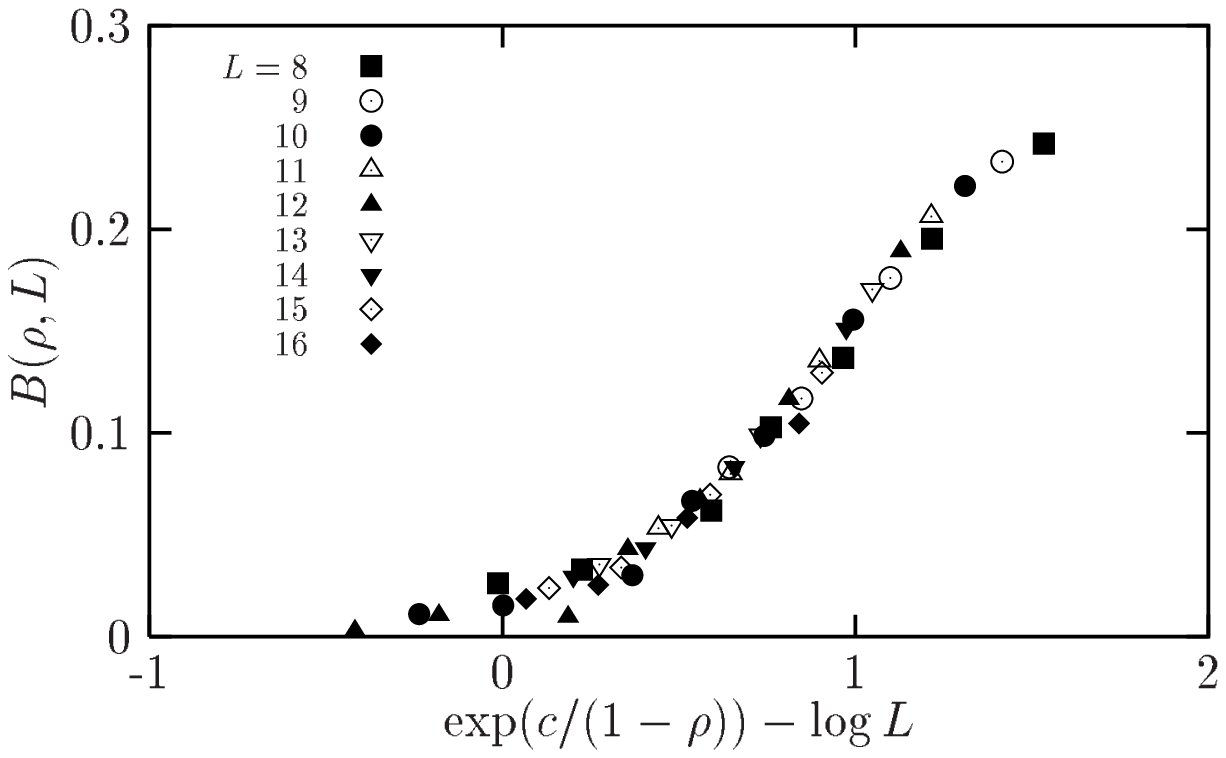,width=8.65cm}
\caption{\label{ka}
The parameter $B(T,L)$ as a function
of the reduced variable $\log ( \ell(\rho) / L )$ with
$\ell = (\rho_c - \rho)^{-\alpha}$ (top) or
$\ell = \exp \exp (c/(1-\rho))$ (bottom).
A much better collapse is obtained in the bottom part, 
with one less free parameter.
Systematic deviations are visible in the top part.
Both are for the kinetically constraint lattice glass of
Ref.~\cite{KA}.}

\end{figure}

We show our results for the determination 
of the critical density in Fig.~\ref{ka} where the 
parameters $B(\rho,L)$ are represented as a function
of two possible scaling variables $\ell/L$. 
The top figure assumes a power law divergence of the coherence length 
$\ell(\rho) \approx (\rho_c -\rho)^{-\alpha}$ at a finite 
density $\rho_c < 1$, while the bottom figure 
makes use of the results of Ref.~\cite{giulio}, with a divergence 
at $\rho_c=1$ only.
The best collapses obtained in the two cases are shown, using the values
$\rho_c=0.882$, $\alpha=0.59$, and $c=0.167$. 
The second scaling is clearly 
superior, the first one having systematic deviations
either at large or small sizes. The figure shown here is a compromise 
between those.
Moreover, the second scaling has one less free parameter.
The same conclusions are drawn from the susceptibility (\ref{sus}).
These three arguments thus discriminate
both possibilities, and we numerically confirm the absence
of a dynamical arrest at finite density in this model.

To summarize, we have shown that finite size scaling
techniques can be used
to study the glass transition. 
This follows from the identification of local two-time
observables $F(r,t)$ as relevant `order parameters' with 
interesting fluctuations and correlations.
We have defined two quantities, $B(T,L)$ and $\chi(T,L)$, whose 
scaling behaviour allows one to locate the glass transition
in a much more accurate manner than an extrapolation
of simpler quantities such as the relaxation time.
The method proposed here straightforwardly applies to any
glass-forming systems, including off-lattice models.
More generally, our results show that 
the glass transition has more in common with continuous phase
transitions than usually believed and raise the hope 
that more concepts and methods borrowed from continuous
phase transitions studies can also be used.

\begin{acknowledgments}
I thank JP Garrahan and W Kob who provided me with their 
continuous time algorithms~\cite{jp,KA}, and G Biroli for 
communicating results before publication~\cite{giulio}. 
Over the last 14 months, daily conversations on this subject
with JP Garrahan were very useful. This work is supported 
by a European Marie Curie Fellowship No HPMF-CT-2002-01927, CNRS (France)
and Worcester College Oxford. Numerical results were obtained 
on Oswell at the Oxford Supercomputing Center, Oxford University. 
\end{acknowledgments}

\end{document}